\documentclass[pra,aps,10pt,twocolumn,superscriptaddress,showpacs,amsmath,amssymb,amsfonts]{revtex4-1}
\usepackage{bm}
\usepackage{graphicx}
\usepackage{subfigure}
\usepackage{array}
\usepackage[usenames,dvipsnames]{color}
\usepackage{natbib}
\usepackage{mathrsfs}
\usepackage[breaklinks=true,pdfborder={0 0 0},colorlinks=true,linkcolor=MidnightBlue,citecolor=MidnightBlue,urlcolor=MidnightBlue]{hyperref}

\def\avgd{\overline{g}_{d}}
\def\avmuj{\overline{\mu}_{j}}

\begin{document}

\title{Faraday patterns in coupled one-dimensional dipolar condensates}
\author{Kazimierz {\L}akomy}
\affiliation{Institut f\"ur Theoretische Physik , Leibniz Universit\"at, 
Hannover, Appelstrasse 2, D-30167, Hannover, Germany}
\author{Rejish Nath}
\affiliation{IQOQI and Institute for Theoretical Physics, 
University of Innsbruck, A-6020 Innsbruck, Austria}
\author{Luis Santos}
\affiliation{Institut f\"ur Theoretische Physik , Leibniz Universit\"at, 
Hannover, Appelstrasse 2, D-30167, Hannover, Germany}
\date{\today}

\begin{abstract}
We study Faraday patterns in quasi-one-dimensional dipolar Bose-Einstein 
condensates with parametrically driven dipolar interactions. We show that 
in the presence of a roton minimum in the excitation spectrum, the emergent 
Faraday waves differ substantially in two- and one-dimensional geometries, 
providing a clear example of the key role of confinement dimensionality in 
dipolar gases. Moreover, Faraday patterns constitute an excellent tool to 
study non-local effects in polar gases, as we illustrate for two parallel 
quasi-one-dimensional dipolar condensates. Non-local interactions between 
the condensates give rise to an excitation spectrum characterized by symmetric 
and anti-symmetric modes, even in the absence of hopping. We show that this 
feature, absent in non-dipolar gases, results in a critical driving frequency 
at which a marked transition occurs between correlated and anti-correlated 
Faraday patterns in the two condensates. Interestingly, at this critical frequency, 
the emergent Faraday pattern stems from a spontaneous symmetry breaking mechanism.
\end{abstract}

\maketitle

\section{Introduction}
Inter-particle interactions play an essential role in the physics of 
ultra-cold gases. Although in many experiments these interactions may 
be approximated by a contact potential, there is a rapidly growing interest 
in a novel type of cold gases in which electric or magnetic dipole-dipole 
interactions (DDI) are crucial for the occuring phenomena. These so-called 
dipolar gases include atoms with large magnetic moments~\cite{Griesmaier:2005,
Lev:2011,Ferlaino:2012}, polar molecules~\cite{Weidemuller:2008,Ni:2008,
Ospelkaus:2010} and Rydberg gases~\cite{Tong:2004}. The distinct nature of 
the dipolar interactions leads to a wealth of novel physics, including a 
geometry-dependent stability and a roton-like minimum in the excitation 
spectrum~(for reviews see e.g.~\cite{Baranov:2008,Lahaye:2009}). 

Interestingly, the long-range dipolar interactions result in an inherent 
non-local nature of dipolar gases, particularly striking in deep optical 
lattices. For non-dipolar systems, gases trapped in different sites of a 
deep lattice do not interact with each other. Hence, for a vanishing 
inter-site hopping, different sites may be considered as independent, 
uncorrelated experiments. In contrast, inter-site DDI play a substantial 
role even in the absence of hopping. Recent lattice experiments have shown 
that the inter-site dipolar interactions are a key element in the 
dynamics~\cite{Fattori:2008}, as well as in the stability and collapse 
of dipolar condensates ~\cite{Muller:2011,Billy:2012}. 

Faraday patterns constitute a paradigmatic example of pattern formation 
in periodically driven systems~\cite{Faraday:1831,Cross:1993} ranging from 
classical fluids \cite{Binks:1997}, through multimode lasers~\cite{Szwaj:1998} 
and superfluid Helium \cite{Abe:2007}. Interestingly, Faraday patterns may 
be observed in BECs by modulating the nonlinearity resulting from the interatomic 
interactions~\cite{Staliunas:2002,Staliunas:2004,Nicolin:2007,Nicolin:2011,NicolinVariational:2012,Nicolin:2012}, 
as shown in recent experiments~\cite{Hoefer:2007}. Faraday patterns in Bose-Einstein 
condensates~(BECs) may be directly linked to the spectrum of elementary excitations, 
and in this sense provide an excellent insight into fundamental properties of 
the condensates. In non-dipolar gases the Faraday pattern selection is determined 
uniquely for each modulation frequency due to the monotonically growing 
character of the excitation energy \cite{Dalfovo:2006}. Interestingly, 
this is not anymore the case for dipolar BECs with a roton-like minimum 
in the excitation spectrum \cite{Santos:2003}. As a result, it has been shown 
that Faraday patterns in two-dimensional (2D) dipolar condensates present 
remarkable qualitative novel features~\cite{Nath:2010}  

In this paper, we analyze quasi-one-dimensional (quasi-1D) dipolar condensates 
with periodically driven dipolar interactions. We demonstrate that Faraday patterns 
provide a clear example of the non-trivial role of confinement dimensionality 
in dipolar gases, showing that in the presence of a roton-like minimum in 
the excitation spectrum, Faraday patterns in a quasi-1D trap differ significantly 
with respect to the 2D case~\cite{Nath:2010}. Moreover, Faraday patterns provide as 
well an excellent tool for a study of non-local effects in dipolar condensates, as 
we illustrate with two parallel quasi-1D BECs, in the absence of tunneling. The 
non-local dipolar interactions between both BECs lead to an unfolding of the 
excitation spectrum into symmetric and anti-symmetric modes with respect to 
the transposition of the two condensates. We show that, as a consequence, at a 
critical driving frequency a transition between correlated (symmetric) and 
anti-correlated (anti-symmetric) Faraday patterns in the two BECs occurs. For 
the critical driving the emergent Faraday pattern differs from one realization 
to another, resulting from a spontaneous symmetry breaking mechanism.

The paper is structured as follows. In Sec.~\ref{sec:Model} we introduce the 
model for periodically driven quasi-1D dipolar condensates. Section~\ref{sec:Single} 
is devoted to Faraday patterns in a single quasi-1D BEC, with a focus on the 
differences as compared to 2D condensates. Section~\ref{sec:Double} is dedicated 
to the effects of the inter-condensate dipolar interactions on the Faraday pattern 
selection in two parallel disjoint quasi-1D dipolar condensates. We conclude in 
Sec.~\ref{sec:Conclusions}.

\section{Model}
\label{sec:Model}

We consider in this paper quasi-1D dipolar BECs, either in a single trap~(Sec. III) 
or in two parallel traps~(Sec. IV). Since the former case may be considered as a 
particular realization of the latter, we present in this Section the general formalism 
for parallel quasi-1D BECs aligned along the $z$ axis, and separated along the $y$ 
axis by a distance $\Delta$. We assume the potential barrier separating both quasi-1D 
BECs sufficiently large to suppress any hopping between them. Each condensate 
experiences a strong harmonic confinement of frequency $\omega_{\scriptscriptstyle \perp}$ 
in the $x$-$y$ plane and no confinement along the $z$ direction. The atoms possess a 
magnetic dipole moment $\mu$~(the results are equally valid for electric dipoles) 
oriented by an external field along the $y$ axis. We employ dimensionless expressions, 
using units of frequency $\omega_{\scriptscriptstyle \perp}$ and length 
$l_{\scriptscriptstyle \perp} =\sqrt{\hbar/M\omega_{\scriptscriptstyle \perp}}$, with 
$M$ the particle mass.

Due to the strong $x$-$y$ confinement we assume that the system remains 
in the ground state of the $x$-$y$ harmonic oscillator (this condition 
is self-consistently verified), and employ the non-local non-linear Schr{\"o}dinger 
formalism developed in Refs.~\cite{Lakomy:20121,Lakomy:20122} for a stack of 
quasi-1D dipolar BECs, to obtain the coupled equations for the wave functions 
$\psi_{j}(z)$ in traps $j=1,2$:
\begin{multline}
 i \partial_{t} \psi_{j} \left( z \right) = \Biggl[ -\frac{1}{2} \partial_{z}^{2} 
 + g n_{j}(z) \\ 
 + \frac{2\pi}{3} g_{d} \sum_{m} \int\! d k_z e^{i k_z z}  \hat{n}_{m} \left( k_z \right) 
 \!F_{|m-j|}\left(k_z \right) \! \Biggr] \psi_{j} \left( z \right).
\label{eqn:1DGPsystem}
\end{multline}
Short-range interactions are characterized by the coupling 
constant $g = g^{3\text{D}} n_{0}/2 \pi \hbar \omega_{\scriptscriptstyle \perp} l^{3}_{\scriptscriptstyle \perp}$, 
where $n_0$ is the linear density, and $g^{3\text{D}} = 4 \pi a_{sc} \hbar^2/M$, 
with $a_{sc}$ the $s$-wave scattering length. The DDI are determined by the coupling constant 
$g_{d} = g^{3\text{D}}_{d} n_{0}/2 \pi \hbar \omega_{\scriptscriptstyle \perp} l^{3}_{\scriptscriptstyle \perp}$, 
where $g^{3\text{D}}_{d} =  \mu_0 \mu^2/4 \pi$, with $\mu_{0}$ the vacuum permeability.
In Eq.~\eqref{eqn:1DGPsystem}, $\hat{n}_{m} \left( k_z \right)$ is the Fourier transform of the 
linear density $n_{m}(z)=\left| \psi_{m} \left( z \right) \right|^{2}$, and 
\begin{align}
F_{p} & \left( k_z \right) = \! \int\limits_{0}^{\infty} \!dk \frac{k e^{-\frac{1}{2}k^{2}}}{k^{2}+k_{z}^{2}} \nonumber \\
\times & \left[ \left(k^{2}-2 k_{z}^{2}\right) J_{0} \left(k \Delta p \right) -3 k^{2} J_{2}\left( k \Delta p  \right) \right],
\end{align}
where $J_{n}(x)$ are the Bessel functions of the first kind. 

In the following we consider a parametric modulation of the dipole-dipole interactions
\begin{equation}
g_{d}(t) = \avgd (1+2 \alpha \cos(2 \omega t)),
\label{eqn:gdModulationFormula}
\end{equation}
where $\alpha$ characterizes the modulation strength. Such modulation may be implemented 
with intensity oscillations of the polarizing electric field for the case of polar molecules, 
or with additional transverse magnetic fields, which lead to a precession of the dipole 
moment orientation, for the case of magnetic dipoles. 

The modulation of $g_d$ induces Faraday waves. With the aim of examining the growth of 
such patterns, we introduce the following ansatz for the wave functions:
\begin{equation}
\psi_{j} \left( z,t \right) = \psi_{j_{\scriptstyle{H}}} \left( 1 + A_{j}(t) \cos(q z) \right),
\label{eqn:dynamicsAnsatz}
\end{equation}
which describes well the physics of the pattern in the linear regime, where the modulation 
is weak and we may consider each momentum component $q$ of the pattern separately. 
In Eq.~\eqref{eqn:dynamicsAnsatz}, we introduce the complex amplitude $A_{j}(t) = u_{j}(t) + i v_{j}(t)$, 
which determines the perturbation from the initial homogeneous solution
$\psi_{j_{\scriptstyle{H}}} = \exp\{ -i \avmuj [ t + (\Omega_{j}/\omega) \sin ( 2 \omega t ) ]\}$, where
$\Omega_{j} = \alpha ( 1- \avgd /\avmuj )$, 
and $\avmuj = g + \frac{2\pi}{3} \avgd \sum F_{|m-j|}(0)$ is the 
chemical potential. Inserting Eqs.~\eqref{eqn:gdModulationFormula} and~\eqref{eqn:dynamicsAnsatz},
into Eq.~\eqref{eqn:1DGPsystem}, and linearizing in $A_{j}$ we arrive at the 
system of equations describing the modulation dynamics
\begin{align}
&\frac{d^{2}u_{j}}{dt^{2}} + \frac{q^2}{2} \Biggl[
\left( \frac{q^2}{2} + 2 g \right) u_{j} \nonumber \\ & \;\;\;\;\;
+ \frac{4\pi}{3} g_{d}(t)  \sum\limits_{m} u_{m} F_{|m-j|}(q)
\Biggr] = 0. 
\label{eqn:mathieuGeneral}
\end{align}
\begin{figure}[t]
\includegraphics[width=1.\columnwidth]{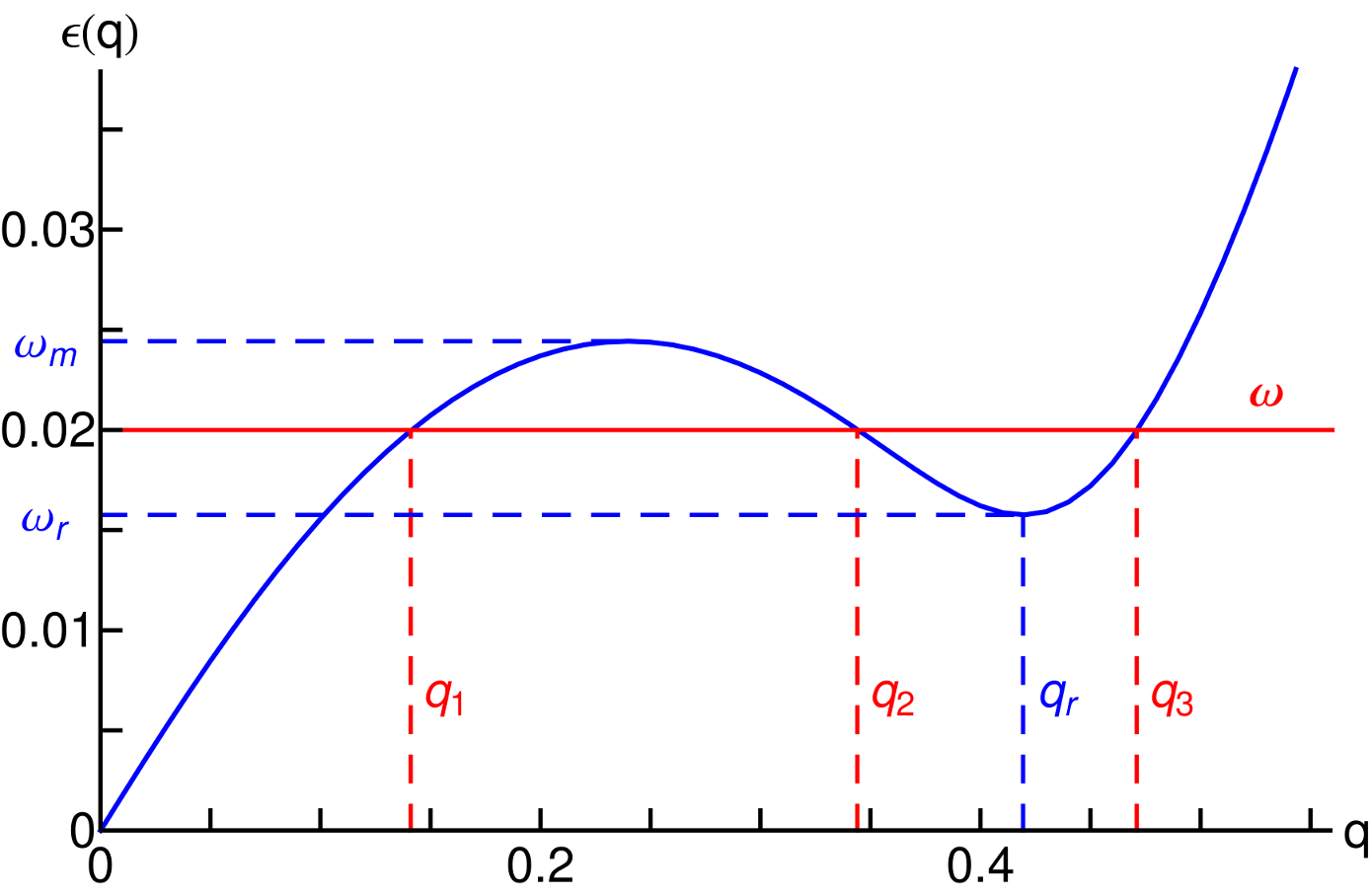}
\caption{(Color online) Excitation spectrum $\epsilon(q)$ 
of a single quasi-1D dipolar BEC, with $g=-0.1007$, 
$g_{d}=0.0629$. Note the roton minimum at $\omega_r=\epsilon(q_r)$ 
and the maxon maximum at $\omega_m$. For a driving 
frequency $\omega_r<\omega<\omega_m$ there are three 
possible momentta $q_{1,2,3}$ obeying the resonance 
condition $\epsilon(q)=\omega$. 
}
\label{fig:1}
\end{figure}

\section{Faraday Patterns in a single quasi-1D dipolar BEC}
\label{sec:Single}

We consider in this Section the case of a single condensate, being 
particularly interested in the differences between the emergent Faraday 
patterns in a quasi-1D trap and those predicted in Ref.~\cite{Nath:2010} 
for a 2D condensate. Employing a similar Bogoliubov analysis as the one 
presented in Ref.~\cite{Lakomy:20121}, we obtain the spectrum of 
elementary excitations in the considered case~(see Fig.~\ref{fig:1}): 
\begin{equation}
\epsilon(q) = \sqrt{ \frac{q^2}{2} \left( \frac{q^2}{2} + 
  2 g + \frac{4\pi}{3} g_d F_0(q) \right) }
\label{eqn:bogolibovSingleTube}
\end{equation}
where 
\begin{equation}
F_{0}(q) = 1 + \frac{3}{2} q^{2} e^{q^2/2} \text{Ei}\left(-q^2/2\right),
\label{eqn:Hdefinition}
\end{equation}
with Ei($x$) the exponential integral function. 
Using Eqs.~\eqref{eqn:mathieuGeneral} and~\eqref{eqn:bogolibovSingleTube}, 
we arrive at the corresponding Mathieu equation~\cite{McLachlan:1951}:
\begin{equation}
\frac{d^{2}u}{dt^{2}} + \left[ \epsilon^{2}(q) + 
  2 \omega^{2} b(q,\omega,\alpha) \cos(2 \omega t) \right] u = 0
\label{eqn:mathieuSingle}
\end{equation}
with
\begin{equation}
b(q,\omega,\alpha) = \frac{2\pi}{3 \omega^{2}} \avgd \alpha q^{2} F_0(q).
\label{eqn:defB}
\end{equation}

Following Floquet theorem~\cite{Morse:1953}, the solutions of 
Eq.~\eqref{eqn:mathieuSingle} are of the form 
$u(t) = e^{\tilde\sigma t}f(t)$ where $f(t)=f(t+\pi/\omega)$ 
and $\tilde\sigma(q,\omega,\alpha)$ is the Floquet characteristic 
exponent, which can be found numerically. If the real part 
$\sigma\equiv\text{Re}(\tilde\sigma)>0$, the homogeneous quasi-1D 
BEC becomes dynamically unstable and Faraday patterns emerge. The typical 
wave length of the pattern will be determined by the most unstable 
mode, i.e., that with the largest $\sigma$. In the limit of small 
driving amplitude, $\alpha \rightarrow 0$, the properties of the pattern 
are governed by momenta $q$ obeying parametric resonances 
$\epsilon_n(q) = n\omega$. 
 
\begin{figure}[t]
\includegraphics[width=1.\columnwidth]{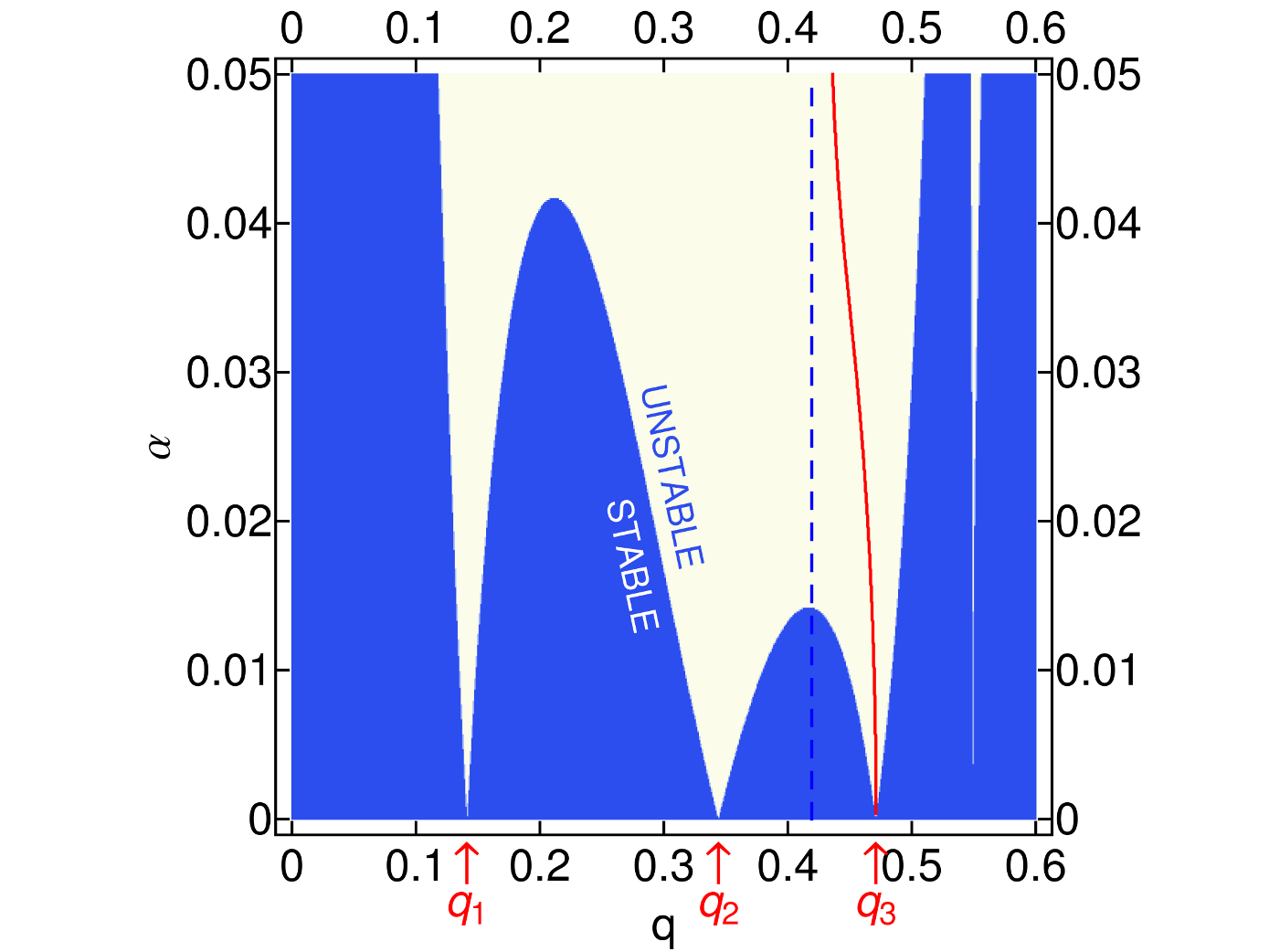}
\caption{(Color online) Stability diagram for the 
parameters of Fig.~\ref{fig:1}, as a function of the 
perturbation strength $\alpha$ and momentum $q$. The 
unstable region is depicted in white. The solid line 
indicates the most unstable mode, and the dashed line 
the roton momentum.}
\label{fig:2}
\end{figure}

Contrary to non-dipolar BECs with a monotonic  spectrum $\epsilon(q)$, 
dipolar gases may offer a more complex roton-maxon 
spectrum~\cite{Santos:2003}~(Fig.~\ref{fig:1}). As a consequence of 
this non-monotonic character, for a specific range of $\omega$, between 
the roton and maxon frequencies~($\omega_r$ and $\omega_m$, respectively) 
there are three values $q_1<q_2<q_3$ satisfying the resonance condition 
$\epsilon(q) = \omega$. Figure~\ref{fig:2} shows the stability diagram for a 
driving frequency in this particular window. As expected, for small 
amplitudes $\alpha$, the three instability tongues (white regions) correspond 
exactly to $q_{1,2,3}$ (Fig.~\ref{fig:1}). This opens an interesting question 
about which of the three modes dominates the pattern formation. For a 2D geometry, 
Ref.~\cite{Nath:2010} showed that when modulating dipole-dipole interactions, 
the most unstable mode corresponds to the intermediate momentum $q_2<q_r$, 
with $q_r$ the roton momentum. Crucially, as we show below, this is not the 
case in a quasi-1D dipolar condensate. This striking contrast between quasi-1D 
and 2D predictions illustrates once more the key role played by the trapping 
geometry in dipolar gases.

\begin{figure}[t]
\includegraphics[width=1.\columnwidth]{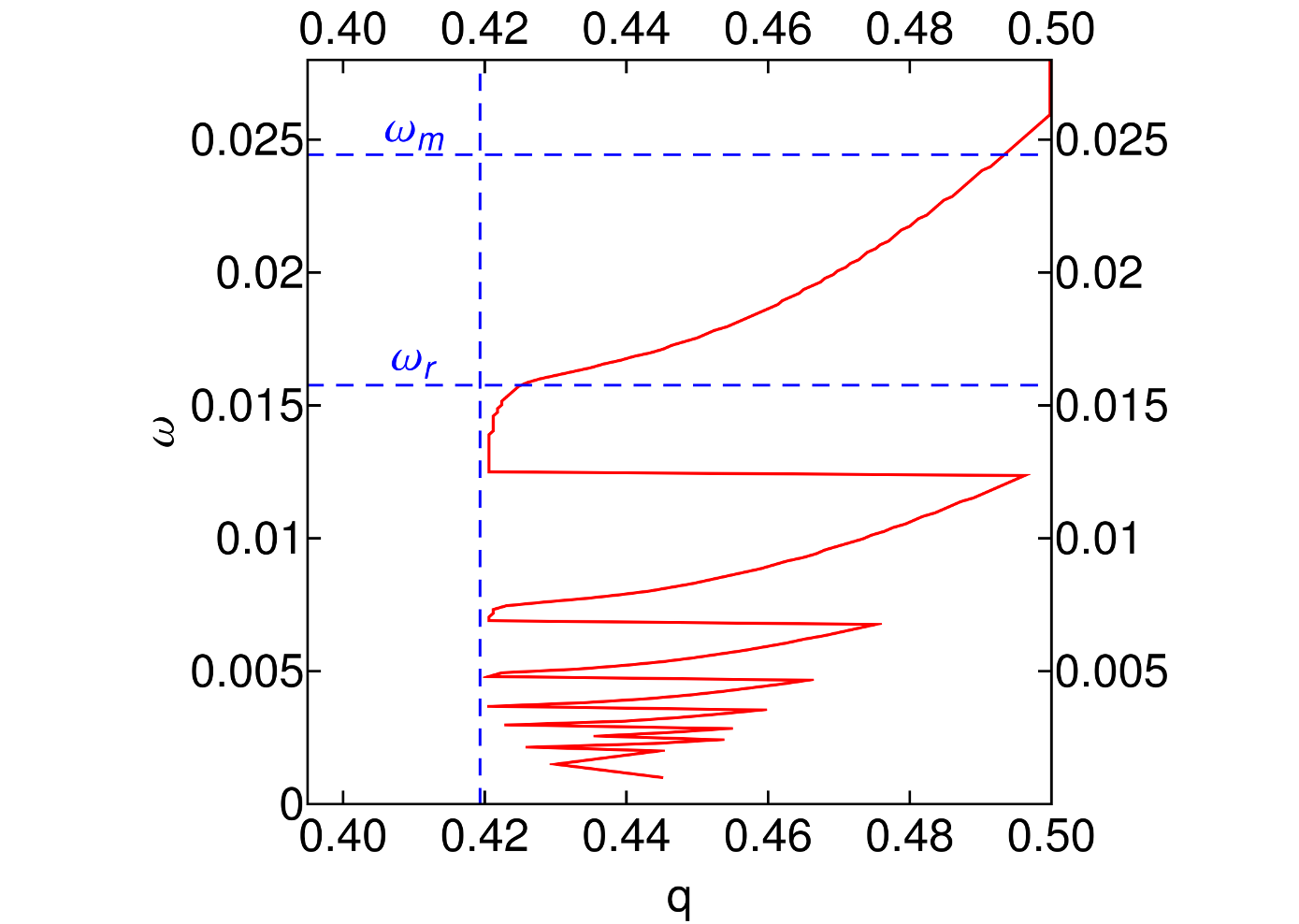}
\caption{(Color online) Most unstable momentum $q$ as 
a function of the driving frequency $\omega$ for the 
parameters of Fig.~\ref{fig:1}, with $\alpha=0.01$. 
The horizontal dashed lines indicate the roton and 
maxon frequencies ($\omega_{r,m}$), and the vertical 
line the roton momentum.}
\label{fig:3}
\end{figure}

The problem of the most unstable mode is best understood employing a 
series expansion of the Floquet exponent with small parameter 
$b(q,\omega,\alpha)$ ~\cite{Tisserand:1894,Nicolin:2007,Nicolin:2008}, 
which, for the first parametric resonance $\epsilon(q) = \omega$, yields 
$\sigma \simeq b(q,\omega,\alpha)/2\propto q^{2} F_0(q)$. Remarkably, 
in contrast to the 2D case, we find that the most unstable mode corresponds 
to the largest momentum $q_3>q_r$~(solid line in Fig.~\ref{fig:2}). 
Fig.~\ref{fig:3} depicts a momentum of the most unstable mode as a function 
of the driving frequency $\omega$. The plot confirms that for all 
frequencies within the window $\omega_m < \omega < \omega_r$ the 
momentum characterizing the most unstable mode is larger than the roton 
momentum, contradicting the prediction for a 2D pancake geometry~\cite{Nath:2010}. 
For $\omega < \omega_r$, alike in the 2D case, the observed modulations 
are dominated by higher resonances with $q$ in the vicinity of $q_r$. 
However, unlike the 2D scenario, even in this regime the most unstable 
mode in a quasi-1D BEC is characterized by $q>q_{r}$. We emphasize that 
the different nature of the Faraday pattern reported here stems solely 
from the quasi-1D character of the condensate, which leads to a specific 
momentum dependence of $b(q,\omega,\alpha)$ that differs from that in 2D.

We have simulated numerically the time evolution of the non-local 
non-linear Schr\"odinger equation~\eqref{eqn:1DGPsystem} with the 
parametrically driven nonlinearity, according to Eq.~\eqref{eqn:gdModulationFormula}. 
The emergent pattern has been examined by means of Fourier transform of 
the condensate density, which confirmed the results for the most 
unstable mode that we obtained within the Mathieu analysis.

\section{Faraday Patterns in two 1D dipolar BEC\lowercase{s}}
\label{sec:Double}

We now turn to the study of Faraday patterns in two parallel 
quasi-1D dipolar BECs. For non-dipolar BECs, in the absence of hopping, 
each BEC behaves independently, and hence an experiment with two 
BECs reduces to two uncorrelated experiments with a single condensate. 
The situation is radically different in dipolar BECs, since, despite 
the absence of hopping, the non-local character of the dipolar potential 
gives rise to a coupling between the two BECs, with the strength of 
the inter-condensate interactions governed by $F_1(k_z)$. These non-local 
interactions lead to a collective character of the elementary excitations 
that are shared among the two quasi-1D condensates~\cite{Klawunn:2009,Lakomy:20121}. 
Consequently, the excitation spectrum unfolds into two branches 
\begin{equation}
\epsilon_{\pm}(q) = \sqrt{ \frac{q^2}{2} \left ( \frac{q^2}{2} + 2 g + \frac{4\pi}{3} g_d 
\left ( F_0(q) \pm F_1(q) \right ) \right) },
\label{eqn:bogolibovDoubleTubes}
\end{equation}
which correspond, respectively, to symmetric and anti-symmetric states 
with respect to the transposition of traps $j=1\leftrightarrow j=2$.

\begin{figure}[t]
\includegraphics[width=1.\columnwidth]{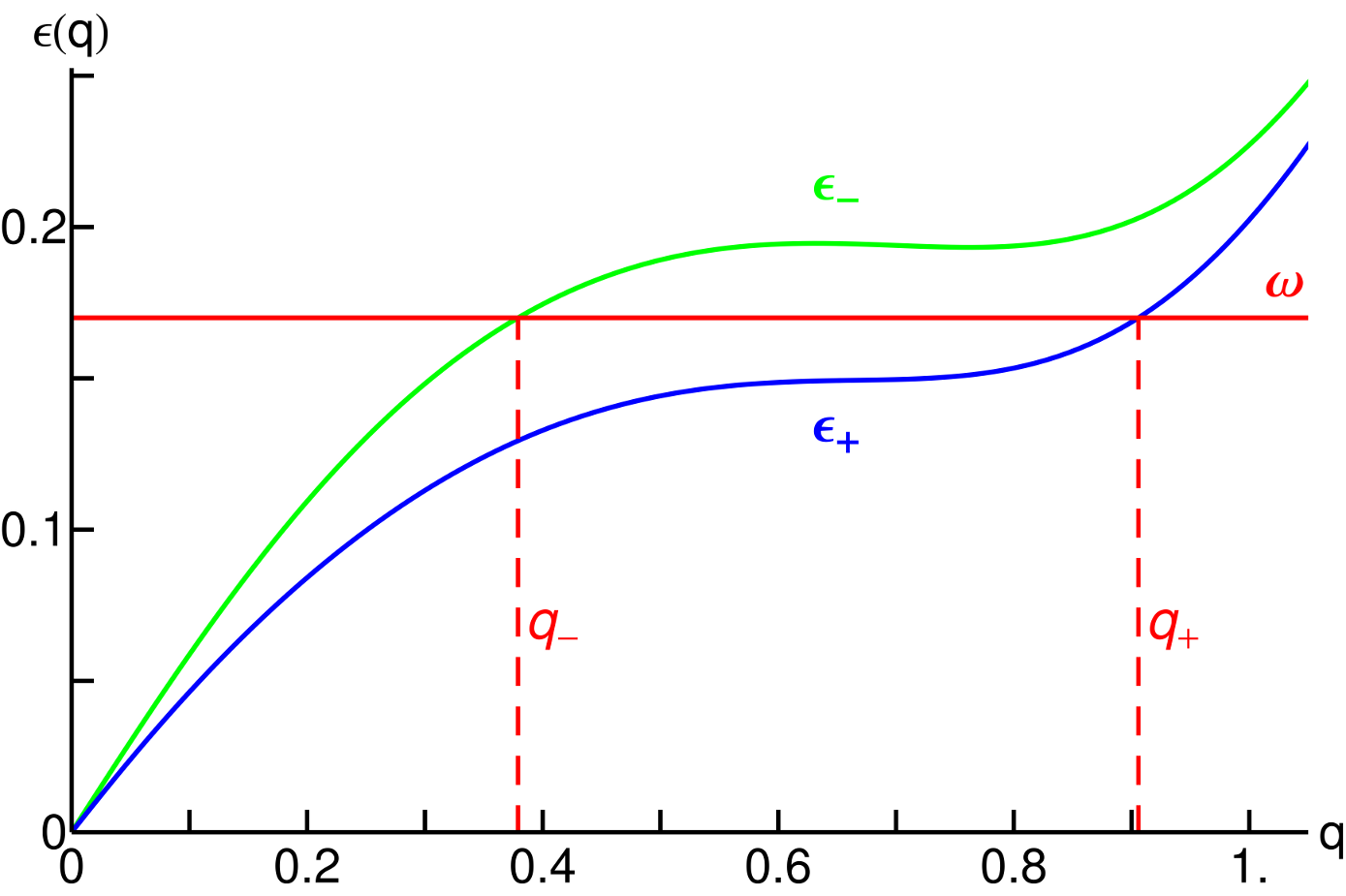}
\caption{(Color online) Elementary excitations of two 
parallel quasi-1D dipolar BECs, for $g=-0.0629$, $\avgd =0.1749$, 
and $\Delta/l_\perp=6$. Note the two branches of the 
elementary excitations $\epsilon_{\pm}(q)$, corresponding, 
respectively, to symmetric and anti-symmetric modes 
with respect to the transposition of traps $j=1\leftrightarrow j=2$. }
\label{fig:4}
\end{figure}
\begin{figure}[b]
\includegraphics[width=1.\columnwidth]{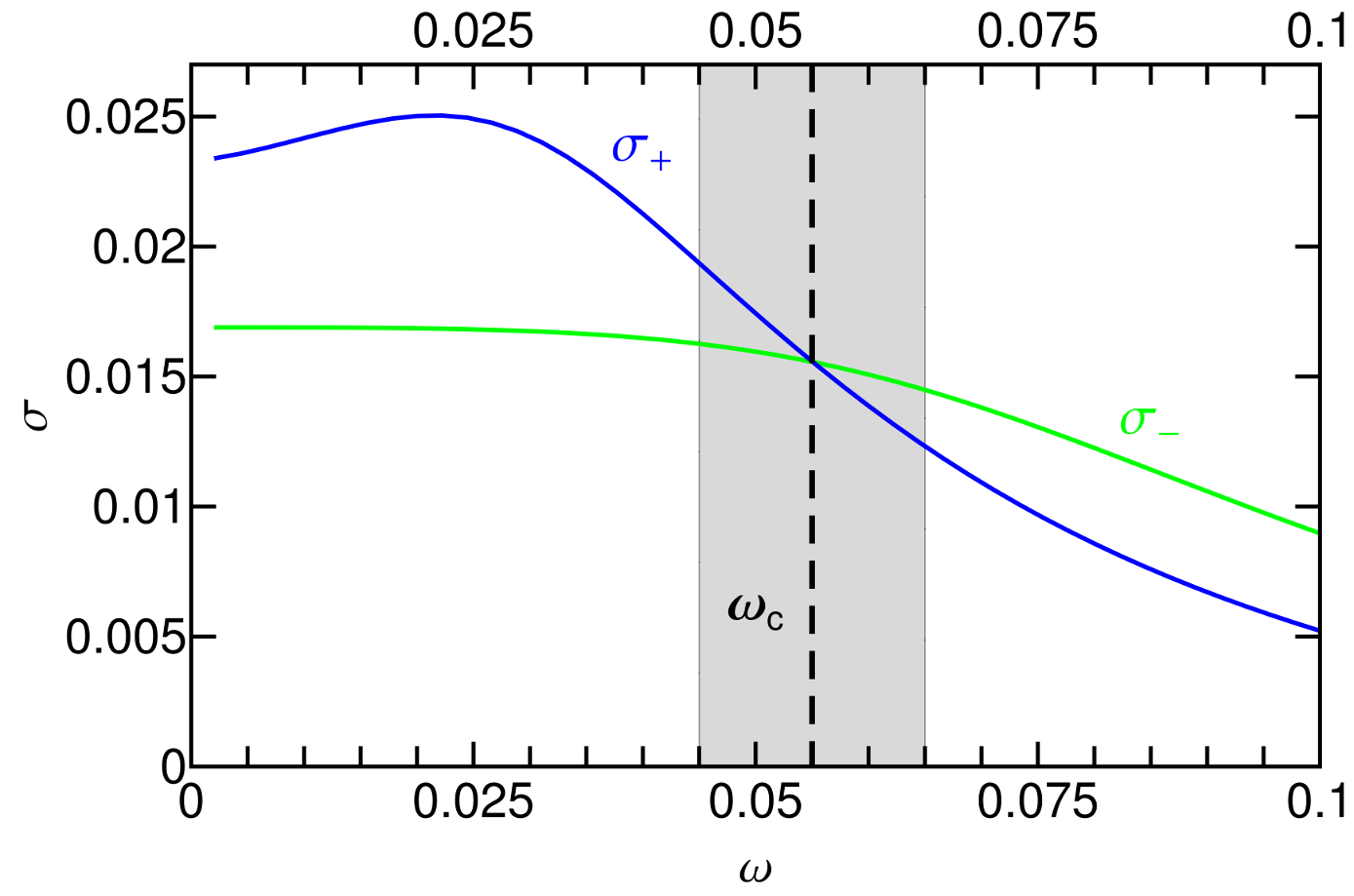}
\caption{(Color online) Real part $\sigma_\pm$ of 
the Floquet exponent, corresponding to the the first 
parametric resonance for the symmetric and anti-symmetric 
excitation branches $\epsilon_{\pm}(q)=\omega$, as a 
function of the driving frequency $\omega$. Note that 
at a critical frequency $\omega_{c}=0.055$, both exponents 
are equal, $\sigma_+=\sigma_-$, indicating a transition 
between the symmetric and the anti-symmetric Faraday pattern. 
In the figure we employ $g=-0.0435$, $\avgd =0.0437$, 
$\Delta=6l_\perp$, and $\alpha=0.02$.}
\label{fig:5}
\end{figure}

Interestingly, this implies that a periodic modulation of the dipolar 
interactions yields two different parametric resonances for each driving 
frequency $\omega=\epsilon_\pm(q_\pm)$, even in the absence of the roton 
minimum~(see Fig.~\ref{fig:4}). Note, that the patterns are characterized 
not only by their momentum $q_\pm$ but also by their symmetric~($+$) or 
anti-symmetric~($-$) character. In analogy to Sec.~\ref{sec:Single}, the double 
solution raises a fundamental question about which of these two modes is the 
most unstable, and hence provides the dominant Faraday pattern.  We stress 
that this non-trivial physics stems directly from the inter-condensate 
interactions, which lead to the splitting between the two branches in the spectrum, 
being a qualitatively new feature of dipolar condensates.

Similarly to the previous section, we employ Eqs.~\eqref{eqn:mathieuGeneral} 
for $j=1,2$, and the spectra~\eqref{eqn:bogolibovDoubleTubes}. In turn, we obtain 
two decoupled Mathieu equations for the symmetric and anti-symmetric combinations 
$u_\pm =u_1\pm u_2$:
\begin{equation}
\frac{d^{2}u_{\pm}}{dt^{2}} + \left[ \epsilon^{2}_{\pm}(q) 
  + 2 \omega^{2} b_{\pm}(q,\omega,\alpha) 
\cos(2 \omega t) \right] u = 0
\label{eqn:mathieuMagic}
\end{equation}
with
\begin{equation}
b_{\pm}(q,\omega,\alpha) = \frac{2\pi}{3 \omega^{2}} \avgd \alpha q^{2} \left ( F_0(q)\pm F_1(q) \right ),
\end{equation}
to which we apply the Floquet analysis employed in the study of 
Eq.~\eqref{eqn:mathieuSingle}. As in the case of a single BEC, the 
first parametric resonances $\epsilon_\pm (q_\pm)=\omega$ are characterized 
by the Floquet exponent $\sigma_\pm \simeq b_{\pm}(q,\omega,\alpha)/2 
\propto q^{2} ( F_0(q)\pm F_1(q) )$, and the emerging Faraday pattern is 
determined, for each driving frequency separately, by the mode with the 
largest $\sigma$. Remarkably, the involved momentum dependence of 
$F_0(q)\pm F_1(q)$ leads to an intricate relation between the Floquet 
exponents and the driving frequency $\omega$, as presented in Fig.~\ref{fig:5}.

\begin{figure}[t]
\includegraphics[width=1.\columnwidth]{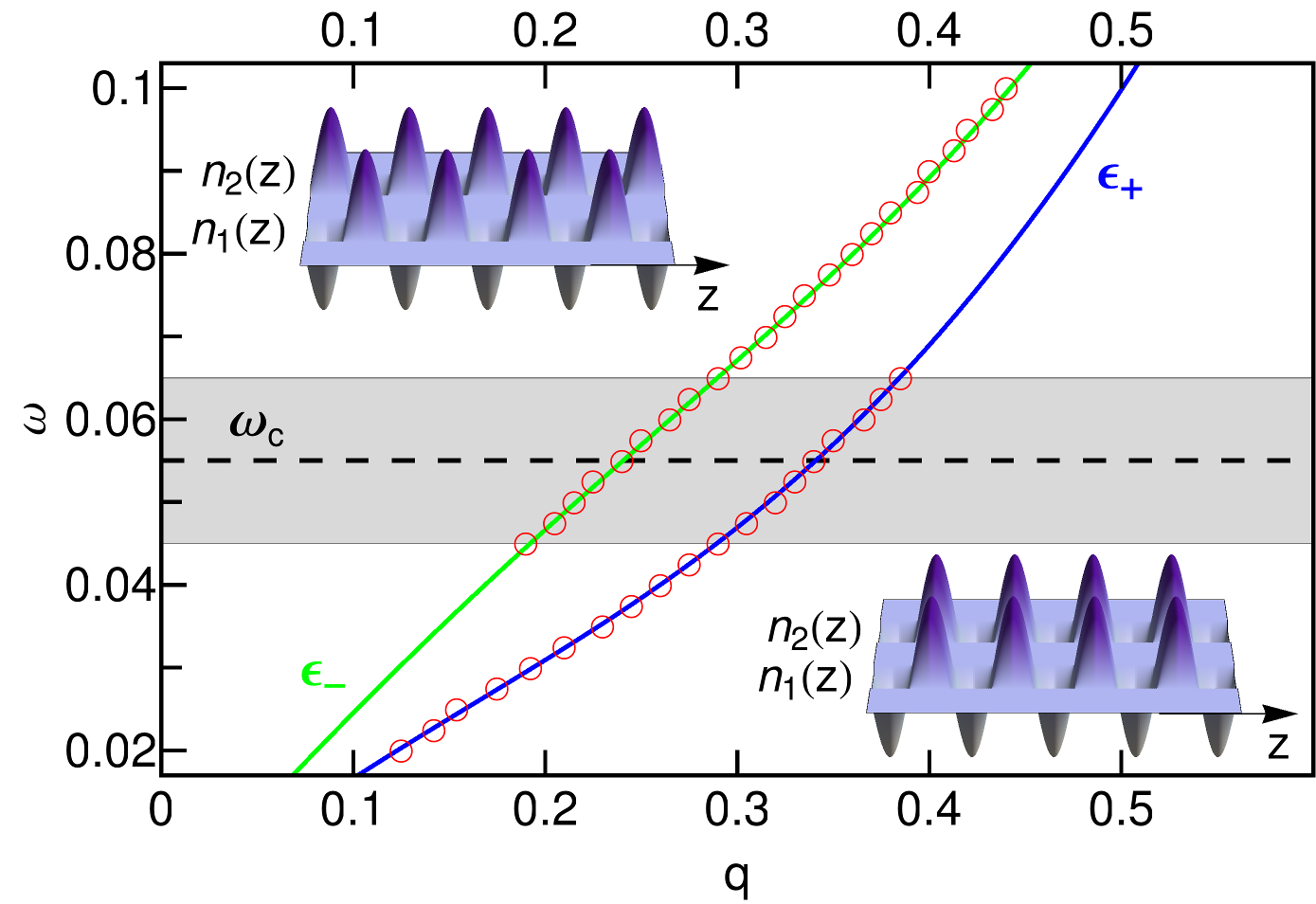}
\caption{(Color online) Analysis of the pattern selection 
as a function of the driving frequency $\omega$, in the 
neighborhood of the critical frequency $\omega_{c}$ 
(for the same parameters as in Fig.~\ref{fig:5}). The 
solid lines represent the excitation branches. For each 
$\omega$ we indicate with a circle a momentum value where 
the numerical Fourier transform $\hat{n}_{j}(k_{z})$ 
of the Faraday pattern shows a clear maximum. For $\omega$ 
well below~(above) $\omega_{c}$  we observe a single peak 
at $q_+$~($q_-$) indicating that a symmetric~(anti-symmetric) 
Faraday pattern emerges~(see insets). In the vicinity of 
$\omega_{c}$ (shaded region), both modes are equally unstable 
and we observe the two corresponding peaks occuring 
simultaneously in the Fourier transform (see text).}
\label{fig:6}
\end{figure}

\begin{figure}[b]
\includegraphics[width=1.\columnwidth]{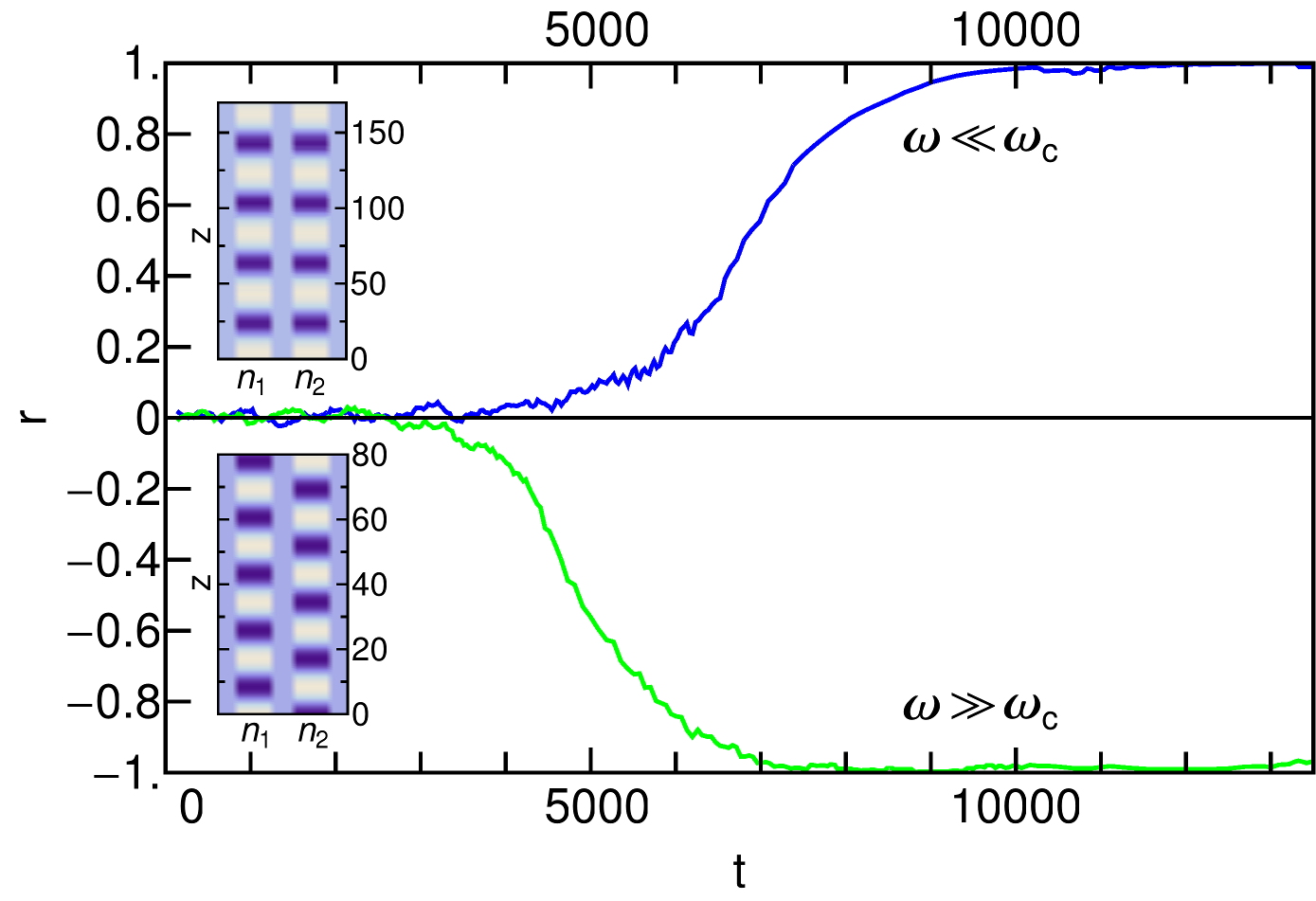}
\caption{(Color online) Correlation function $r(t)$ 
for the same parameters as in Fig.~\ref{fig:6}~($\omega_{c}=0.055$).
The upper curve, which corresponds to $\omega=0.025<\omega_{c}$, 
approaches $r=1$ indicating a perfectly correlated pattern 
in both quasi-1D traps. The lower curve, which corresponds 
to $\omega=0.08>\omega_{c}$, reaches $r=-1$ proving a perfect 
anti-correlation between the Faraday patterns in the two traps. 
The insets show the corresponding numerical results for the 
density distribution $n_j(z)$ for $t=11000$, with the bright 
(dark) colors indicating density maxima (minima). Naturally, 
for sufficiently long times, well beyond the linear regime, 
the Faraday patterns and their correlations are eventually 
destroyed.}
\label{fig:7}
\end{figure}

Crucially, the curves $\sigma_\pm (\omega)$ cross at a critical 
frequency $\omega_{c}$. In consequence, we expect a distinct transition, 
as a function of the driving frequency $\omega$, between the symmetric Faraday 
pattern for $\omega<\omega_{c}$ and the anti-symmetric pattern for 
$\omega>\omega_{c}$. Such transition is marked by an abrupt change of 
the patterns from a maximum-maximum alignement~(correlated patterns) 
to a maximum-minimum alignement~(anti-correlated patterns), as depicted 
in the corresponding insets of Figs.~\ref{fig:6}~and~\ref{fig:7}. 
Moreover, for $\omega=\omega_{c}$ the patterns in both condensates 
exhibit a pronounced change of the wavelength of the modulation, 
from $1/q_+(\omega_{c})$ to $1/q_-(\omega_{c})$. 

This transition has been confirmed by means of direct numerical simulations of 
Eqs.~\eqref{eqn:1DGPsystem}, with the parametric driving governed by 
Eq.~\eqref{eqn:gdModulationFormula}. As for a single condensate, we 
Fourier transform the density of each condensate to obtain the dominant 
momenta of the emergent Faraday patterns. The results, in the vicinity 
of the critical frequency $\omega_{c}$, are depicted in Fig.~\ref{fig:6}, 
where, on top of the spectra $\epsilon_\pm$, for each driving frequency 
$\omega$ we indicate with a circle the momentum value where the numerically 
evaluated $\hat{n}_{j}(k_{z})$ shows a marked maximum. We find that, in 
agreement with the results for $\sigma_\pm(\omega)$ presented in 
Fig.~\ref{fig:5}, for $\omega$ well below $\omega_{c}$ the pattern 
presents a single momentum component at $q_+$, being characterized by a 
correlation between the patterns in both quasi-1D BECs. In contrast, for $\omega$ 
well above $\omega_{c}$ a single momentum component $q_{-}$ is observed, 
and the patterns in the two quasi-1D BECs are anti-correlated.

In order to quantify the transition between correlated and 
anti-correlated patterns we introduce the correlation coefficient
\begin{equation}
 r = \frac{\int \! dz \, S_{n_{1}}(z) \cdot S_{n_{2}}(z)}
 {\sqrt{\int \! dz \, S^{2}_{n_{1}}(z)} \cdot \sqrt{\int \! dz \, S^{2}_{n_{2}}(z)}},
\end{equation}
where $S_{n_{j}}(z) = n_{j}(z)-\overline{n_{j}}$, with $\overline{n_{j}}$ 
the average density in a trap $j$. Pattern correlation is then characterized 
by $r>0$, whereas anti-correlation leads to $r<0$. Fig.~\ref{fig:7} illustrates 
the radically different time evolution of the correlation coefficient 
below and above the critical driving $\omega_{c}$. Clearly, for frequencies 
sufficiently smaller (larger) than $\omega_{c}$ the system arrives at 
perfectly correlated (anti-correlated) pattern with $|r|=1$.

An interesting scenario occurs for driving frequencies in the 
vicinity of the critical $\omega_{c}$~(shaded region in Figs.~\ref{fig:5} 
and ~\ref{fig:6}), where both the symmetric pattern with wavelength 
$1/q_+(\omega_{c})$ and the anti-symmetric pattern with wavelength 
$1/q_-(\omega_{c})$ are equally unstable. As a result, the Fourier 
transform of the density in each quasi-1D BEC shows a simultaneous 
appearance of both momentum peaks, $q_{+}$ and $q_{-}$~(see Fig.~\ref{fig:6}).

\begin{figure}[t]
\includegraphics[width=1.\columnwidth]{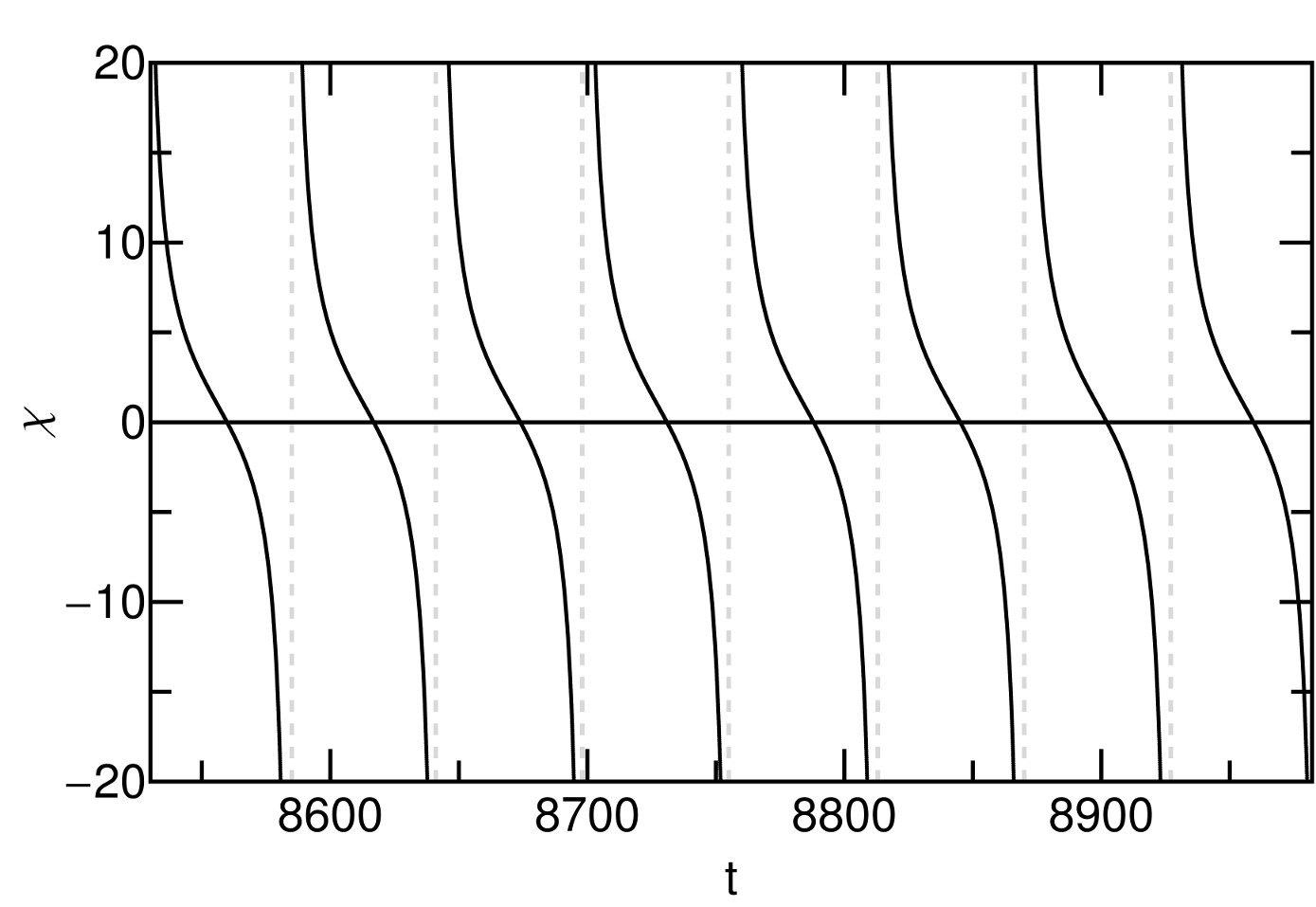}
\caption{(Color online) Population imbalance $\chi(t)$ 
between the two peaks at $q_\pm(\omega_{c})$ for 
the critical driving $\omega=\omega_{c}$ (for the same 
parameters as Fig.~\ref{fig:6}). Note the $2\omega$ periodicity 
($T=\pi/\omega=57.1$) that stems from a spontaneous symmetry 
breaking mechanism (see text).}
\label{fig:8}
\end{figure}

Note that at $\omega=\omega_{c}$, not only $\epsilon_+(q_+)=\epsilon_-(q_-)$ 
but also $b_+(q_+,\omega,\alpha)=b_-(q_-,\omega,\alpha)$ and hence the 
two Mathieu equations \eqref{eqn:mathieuMagic} for $u_+$ and $u_-$ become 
identical. This symmetry is however spontaneously broken in experiments 
due to quantum and thermal fluctuations, which lead to different initial 
conditions (populations) for both modes, that change randomly from 
one realization to another. This spontaneous symmetry breaking mechanism 
is best studied quantitatively by considering the relative weight of the 
momentum peaks at $q_+$ and $q_-$ in the Fourier-transform of the density 
$\hat n(k_z)$. To this end, we define the imbalance parameter 
\begin{equation}
\chi(t) = \frac{\hat n(q_+,t)-\hat n(q_-,t)}{\hat n(q_+,t)+\hat n(q_-,t)}.
\end{equation} 
For $\omega$ well below or above $\omega_{c}$, once the pattern emerges, 
$\chi(t)=\pm 1$. In the vicinity of $\omega_{c}$, however, the imbalance 
parameter $\chi(t)$ shows a clear periodicity with frequency 
$2\omega$~(see Fig.~\ref{fig:8}). Note that these oscillations do not 
result from non-linear competition, as they occur well within the linear regime.
In fact, the $2\omega$ oscillations of $\chi(t)$ originate in different, 
spontanously chosen, initial conditions for $u_{+}$ and $u_{-}$, 
which lead to their different time evolution that can be well 
approximated by $u_{\pm}(t) = (u^{c}_{\pm} \cos(\omega t) 
+ u^{s}_{\pm} \sin(\omega t)) \exp(\sigma \omega t)$, where 
$u^{c\scriptscriptstyle{/}\scriptstyle{s}}_{\pm}$ are the constants 
determined by the initial conditions. Furthermore, spontaneous symmetry-breaking 
leads to a different result for the imbalance $\chi(t)$ from one 
realization to another, what we have confirmed by considering small random 
differences in the initial conditions for our numerical simulations 
of Eqs.~\eqref{eqn:1DGPsystem}.

\section{Conclusions}
\label{sec:Conclusions}

Faraday patterns in dipolar BECs are crucially dependent on the unique 
properties of the dipole-dipole interactions. In particular, due to the 
long-range anisotropic nature of the dipolar interactions, the character 
of the Faraday patterns depends strongly on the dimensionality of 
the condensates. We have shown that for periodically modulated dipolar 
interactions, Faraday patterns in 2D and 1D geometries differ substantially 
in the presence of a roton minimum in the excitation spectrum. Moreover, 
for parallel quasi-1D dipolar BECs, the inter-condensate interactions lead, 
even in the absence of hopping, to an excitation spectrum characterized 
by symmetric and anti-symmetric modes. This, in turn, gives rise  at a 
critical driving frequency to a marked transition between correlated and 
anti-correlated Faraday patterns in the two condensates. Interestingly, at 
this transition point the Faraday pattern selection stems from a 
spontaneous symmetry breaking mechanism.
\section{Acknowledgement}
\label{sec:acknow}
We acknowledge funding by the German-Israeli Foundation, the Cluster of Excellence QUEST and the DFG~(SA1031/6). 

\bibliographystyle{apsrev4-1}
\bibliography{CoupledFaradayPatterns}{}

\end{document}